\documentclass[12pt]{iopart}

\usepackage{iopams}  
\usepackage{perpage}
\usepackage{color}
\usepackage{graphicx}
\usepackage{framed}
\newcommand{\tar}{\tilde \alpha_r}
\newcommand{\bea}{\begin{eqnarray}}
\newcommand{\eea}{\end{eqnarray}}
\newcommand{\beq}{\begin{eqnarray}}
\newcommand{\eeq}{\end{eqnarray}}
\newcommand{\pd}{\partial}
\newcommand{\nd}{\nabla}
\newcommand{\ba}{\begin{eqnarray}}
\newcommand{\ea}{\end{eqnarray}}
\newcommand{\dfrac}{\displaystyle\frac}
\newcommand{\nn}{\nonumber\\}

\newcommand{\be}{\begin{equation}}
\newcommand{\ee}{\end{equation}}
\newcommand{\mc}{\mathcal}
\newcommand{\tx}{\mathrm}

\begin{document}

\title[Black holes and stars in Horndeski theory]{Black holes and stars in Horndeski theory}

\author{Eugeny Babichev$^1$, Christos Charmousis$^1$ and Antoine Leh\'ebel$^1$}

\address{$^1$ Laboratoire de Physique Th\'eorique, CNRS, Univ. Paris-Sud, \\ Universit\'e Paris-Saclay, 91405 Orsay, France.}

\eads{\mailto{eugeny.babichev@th.u-psud.fr}, \mailto{christos.charmousis@th.u-psud.fr}, \mailto{antoine.lehebel@th.u-psud.fr}}

\begin{abstract}
We review black hole and star solutions for Horndeski theory. For non-shift symmetric theories, black holes involve a Kaluza-Klein reduction of higher dimensional Lovelock solutions. On the other hand, for shift symmetric theories of Horndeski and beyond Horndeski, black holes involve two classes of solutions: those that include, at the level of the action,  a linear coupling to the Gauss-Bonnet term and those that involve time dependence in the galileon field. We analyze the latter class in detail for a specific subclass of Horndeski theory, discussing the general solution of a static and spherically symmetric spacetime. We then discuss stability issues, slowly rotating solutions as well as black holes coupled to matter. The latter case involves a conformally coupled scalar field as well as an electromagnetic field and the (primary) hair black holes thus obtained. We review and discuss the recent results on neutron stars in Horndeski theories.
\end{abstract}

\maketitle

\section{Introduction-Horndeski theory}

The prototype modification of General Relativity (GR) is scalar-tensor theory and as such it was introduced a long time ago as a potential gravity theory not obeying the strong equivalence principle \cite{chaBD}. Scalar-tensor theories are most probably the simplest, consistent and non trivial modification of GR \cite{chafarese}. They acquire one additional degree of freedom, represented by a real scalar field and include or contain in some limit many other theories of modified gravity. Examples of the former case are $f(R)$ theories which are very particular scalar-tensor theories, while massive gravity or braneworlds constitute examples that are scalar-tensor theories in some specific limit. Illustrative examples of the latter are the Randall-Sundrum two brane world model \cite{Randall:1999ee} where to lowest order, the inter-brane distance-radion plays the role of the scalar \cite{Garriga:1999yh}. Or, again, the decoupling limit of DGP \cite{Dvali:2000hr} in which a special covariant galileon term appears simulating higher dimensional dynamics \cite{Nicolis:2004qq}. Therefore, scalar-tensor theories constitute a generic theoretical setting where one can test deviations from GR and hopefully soon enough, conduct a plethora of strong gravity observational tests. In particular, strong gravity solutions of such theories are very important due to the incoming data we know now to expect from gravitational waves \cite{gw}. Theoretically, we would like to understand in depth fundamental questions such as, do non-trivial (scalar) black hole solutions exist in scalar-tensor theories, are they physically stable and on what time scales? When do no-hair theorems break down and allow for such solutions? What are the gravity modification effects of such solutions? How close are they to GR black holes, what are their novel features that one might observe via gravitational wave detection? Similar types of questions apply for neutron stars: their existence and stability; what is their mass to radius ratio?
The first signal for a binary black hole merger did not show any deviation from GR; however, the signal itself has at least two surprising features : the coalescing black holes are of surprisingly large size and the resulting merger has relatively slow rotation. It is probable that near future gravitational wave signals will give even more unresolved questions and novel phenomena. 

Theoretically, for black holes, if the scalar field is frozen as one expects from no-hair arguments \cite{Herdeiro:2015waa}, then background solutions are identical to GR and differences would have to be seen at the perturbative level. But if genuine hairy solutions do exist theoretically and are mathematically consistent, one may expect clear observational differences from GR black holes. Similar deviations from GR can be expected for neutron stars; for example deviations due to different mass to radius ratio and quasi-normal modes, or again possible gravitational and scalar radiation from binaries (in the case of scalar-tensor) etc. It is therefore very important to study compact strongly gravitating objects in scalar-tensor theories, the prototype modification of gravity. 

But for a start, we should fix the scalar-tensor theory we wish to study in order to seek hairy or strongly gravitating solutions, as it is known that standard Brans-Dicke gravity does not allow for non GR black holes and cannot screen locally the scalar field. In this article, we will concentrate on higher order scalar-tensor theories, for as we will see, they provide an elegant way to bifurcate no-hair theorems and also present local screening features which are essential for the theory to pass weak gravity tests \cite{Babichev:2013usa}. 

Following the work of Lovelock in 1971 \cite{lovelock}, who established the most general metric theory to acquire second order field equations in an arbitrary number of dimensions,
Horndeski in 1974 \cite{Horndeski:1974wa}, posed and answered the following important question: what is the most general scalar-tensor theory in 4-dimensional spacetime yielding second order field equations?
Let us consider a theory ${\cal L_H}$ involving a real scalar $\phi$ and a metric tensor $g_{\mu\nu}$ endowed with a Levi-Civita connection and a locally regular Lorentzian manifold of spacetime. Consider that the theory in question depends only on these two fields and an arbitrary number of their derivatives,
$$
\label{chahorni}
{\cal L_H}=
{\cal L_H}(g_{\mu\nu}, g_{\mu\nu,i_1},...,g_{\mu\nu,i_1...i_p},\phi,\phi_{,i_1},...,\phi_{,i_1...i_q}) ,
$$
 with $p,q\geq 2$. The finite number of derivatives signifies that we have a finite number of degrees of freedom and hence an a priori effective theory of gravity.
Horndeski required that the theory has second order field equations. The resulting theory proposed  avoids Ostrogradski  instabilities \cite{ostro} and is a priori eligible to have ghost free vacua. 
Note that the requirement to have up to two second derivatives is a sufficient condition to avoid the Ostrorgadski instability, but not necessary. Indeed, recent works~\cite{bhorn} have  shown that the presence of higher than second derivatives does not always lead to theories that are plagued by ghosts. The key is to note that certain higher order gravity theories can be degenerate and thus evade additional ghost degrees of freedom; in other words, they acquire the same number of degrees of freedom as Horndeski theory. This puts them on the same footing as Horndeski theories, thus as a priori healthy theories. Although we will not refer to these theories in detail, we will comment on some of their hairy solutions later on. 

In its modern reformulation, Horndeski theory is written as a generalized galileon Lagrangian,
\begin{equation}
\label{ssgg}
\eqalign{ 
\mc{L} &= \mc{L}_2+\mc{L}_3+\mc{L}_4+\mc{L}_5 ,
\\
\mc{L}_2 &= G_2 ,
\\ 
\mc{L}_3 &= -G_3 \Box \phi ,
\\
\mc{L}_4 &= G_4 R + G_{4X} \left[ (\Box \phi)^2 -(\nabla_\mu\nabla_\nu\phi)^2 \right] ,
\\
\mc{L}_5 &= G_5 G_{\mu\nu}\nabla^\mu \nabla^\nu \phi - \frac{1}{6} G_{5X} \big[ (\Box \phi)^3 - 3\Box \phi(\nabla_\mu\nabla_\nu\phi)^2 \\
&~~~ + 2(\nabla_\mu\nabla_\nu\phi)^3 \big] ,
}
\end{equation}
where $G_2$, $G_3$, $G_4$, $G_5$ are arbitrary functions of $\phi$ and $X=- \pd^\mu \phi \pd_\mu \phi/2$, the canonical kinetic term. Additionally,  in our notation, $f_X$ stands for $\pd f(X)/\pd X$, $R$ is the Ricci scalar, $G_{\mu\nu}$ is the Einstein tensor, $(\nabla_\mu\nabla_\nu\phi)^2 = \nabla_\mu\nabla_\nu\phi \nabla^\nu\nabla^\mu\phi$ and $(\nabla_\mu\nabla_\nu\phi)^3=\nabla_\mu\nabla_\nu\phi \nabla^\nu\nabla^\rho\phi \nabla_\rho\nabla^\mu\phi$. The Horndeski terms are also called generalized galileons.  The scalar field (or galileon) has the property of admitting a special symmetry in flat (nondynamical) spacetime for $G_2\sim G_3 \sim X$ and $G_4\sim G_5 \sim X^2$ , which resembles the Galilean symmetry, hence the name galileon~\cite{Nicolis1}. 
Galileon symmetry is broken for a curved background and for general choice of $G_i$. 
The term ``generalized'' refers to the fact that the functions $G_i$ are arbitrary~\cite{deffayet}, in contrast to ``covariant'' galileon with fixed $G_i$. 
It is important to note that generalized galileon theory and Horndeski theory do not start from the same principle; they turn out however to be identical \cite{jap1}.  

A non-trivial subset of Horndeski theory is that of Fab 4 \cite{Charmousis:2011ea}. This particular theory is the unique subset of Horndeski theory that has the following defining property: it admits ``self-tuning solutions" to an arbitrary bulk cosmological constant for flat spacetime.  The scalar galileon adjusts itself dynamically (via an integration constant that is independent of the action couplings) to accommodate an arbitrary cosmological constant in the bulk action with spacetime being locally flat{\footnote{Different subsets of Horndeski theory exhibit similar self-tuning properties towards a de Sitter attractor rather than a Minkowski one, see \cite{stds}.}}. Fab 4 terms are given by the following four geometric scalars:
\bea
\label{chaeq:john}
{\cal L}_\mathrm{John} &=& \sqrt{-g} \: V_\mathrm{John}(\phi)G^{\mu\nu} \nabla_\mu\phi \nabla_\nu \phi ,\\
\label{eq:paul}
{\cal L}_\mathrm{Paul} &=&\sqrt{-g}\: V_\mathrm{Paul}(\phi)   P^{\mu\nu\alpha \beta} \nabla_\mu \phi \nabla_\alpha \phi \nabla_\nu \nabla_\beta \phi ,\nonumber\\
\label{eq:george}
{\cal L}_\mathrm{George} &=&\sqrt{-g}\: V_\mathrm{George}(\phi) R ,\nonumber\\
\label{eq:ringo}
{\cal L}_\mathrm{Ringo} &=& \sqrt{-g}\: V_\mathrm{Ringo}(\phi) \hat G\nonumber
,\eea
where $P_{\mu\nu\alpha \beta}$ is the double dual of the Riemann tensor:
$$
P^{\mu\nu}{}_{\rho\sigma} =(^\star R^\star)^{\mu\nu}{}_{\rho\sigma} \doteq -\frac{1}{2} \epsilon^{\rho\sigma\lambda\kappa}\,R_{\lambda \kappa}{ }^{\xi \tau} \, \frac{1}{2} \epsilon_{\xi \tau\mu\nu},
%\label{chaddual2}
$$
with $\epsilon_{\mu\nu\rho\sigma}$  the rank 4 Levi-Civita tensor. Furthermore, 
\be
\label{GB}
\hat G=R^{\mu\nu \alpha \beta} R_{\mu\nu \alpha \beta}-4R^{\mu\nu}R_{\mu\nu}+R^2
\ee
is the Gauss-Bonnet scalar. We will see in a forthcoming section that all of these terms with particular potentials appear in the Kaluza-Klein reduction of higher order Lovelock terms \cite{VanAcoleyen:2011mj}. This theory presents several interesting aspects beyond self tuning and we will encounter Fab 4 terms throughout our analysis. They represent interactions between the scalar and very specific spacetime curvature tensors. 
Two of these --- the Einstein Hilbert term and the Gauss Bonnet term --- are Lovelock densities (see for example \cite{aegean}). Moreover, the Einstein and the double dual are the unique rank 2 and 4 tensors which are divergence free. These accumulated properties keep the order of the field equations down and explain their origin from Kaluza-Klein reduction. 
In fact Fab 4 terms are connected to shift symmetric galileons  for Horndeski and beyond Horndeski theory. To see this, let us now move on to shift symmetric galileons before briefly commenting on the beyond Horndeski extension. 

In this article, we will concentrate mostly on Horndeski theories that admit a remnant of galileon symmetry in flat spacetime, namely a shift symmetry for the scalar field $\phi \rightarrow \phi+c$ (with $c$ arbitrary real constant) in arbitrary curved spacetime. For this symmetry, it suffices to consider that $G_2$, $G_3$, $G_4$, $G_5$ are arbitrary functions of $X$ only. Shift symmetry is associated with a Noether current which takes the form
\bea
J^\mu 
 &= -\pd^\mu\phi \bigg( G_{2X} - G_{3X} \Box \phi +G_{4X} R + G_{4XX} \left[ (\Box \phi)^2 -(\nabla_\rho\nabla_\sigma\phi)^2 \right]  
\nn
 &~~~~ +G_{5X} G^{\rho\sigma}\nabla_{\rho}\nabla_{\sigma}\phi -\frac{G_{5XX}}{6} \left[ (\Box \phi)^3 - 3\Box \phi(\nabla_\rho\nabla_\sigma\phi)^2 \right.
\nn
 &~~~~ \left. + 2(\nabla_\rho\nabla_\sigma\phi)^3 \right] \bigg) -\pd^\nu X \bigg( - \delta^\mu_\nu G_{3X} + 2 G_{4XX} (\Box \phi \delta^\mu_\nu - \nabla^\mu\nabla_\nu \phi)
 \nn
 &~~~~  +  G_{5X} G^\mu{}_\nu -\frac12 G_{5XX} \big[ \delta^{\mu}_{\nu}(\Box\phi)^2 - \delta^{\mu}_{\nu}(\nd_\rho\nd_\sigma\phi)^2 -2\Box\phi \nd^\mu\nd_\nu\phi 
 \nn
 &~~~~ +2 \nd^\mu \nd_\rho \phi \nd^\rho \nd_\nu \phi \big]   \bigg) +2G_{4X} R^{\mu}{}_{\rho} \nd^\rho \phi + G_{5X} \bigg( -\Box \phi R^\mu{}_\rho \nd^\rho\phi  
 \nn
 &~~~~ + R_{\rho\nu}{}^{\sigma\mu} \nd^\rho\nd_\sigma\phi \nd^\nu\phi + R_\rho{}^\sigma \nd^\rho\phi  \nd^\mu\nabla_\sigma\phi \bigg) .
\label{covcur}
\eea
Although the identification of Fab 4 terms with the Horndeski functionals $G_i$ is not immediate \cite{Charmousis:2011ea}, clearly the Fab 4 terms are shift symmetric galileons if and only if the Fab 4 potentials are constants. For example, the John term (\ref{chaeq:john}) is obtained setting $G_4 = 1 + \beta X$ and $G_2 = -2 \Lambda+ 2 \eta X$, while $V_\mathrm{John}$ is an arbitrary constant. The Gauss-Bonnet term (or Ringo term above) is a notable exception in that it has shift symmetry if and only if it is coupled to a linear scalar field i.e., $V_\mathrm{Ringo}\sim \phi$. This is because the Gauss-Bonnet term is a topological invariant in 4 dimensions and  can be obtained by the choice $G_2=G_3=G_4=0$, $G_5\sim \ln X$.
We will encounter this term when seeking for theories with hairy black holes. Note finally that the Fab 4 potentials are not in the Horndeski class if they depend on $X$ (unlike the $G_i$'s). $X$-dependent Fab 4 potentials will inevitably lead to higher order field equations. It is intriguing however, that certain beyond Horndeski Lagrangians involve Fab 4 terms John and Paul with potentials that are precisely $X$ dependent \cite{bhorn,Babichev:2015qma}. We will mention solutions to these theories when discussing black holes of shift symmetric theories. 
But first, we will show briefly the relation of Horndeski theory to Lovelock and how one can obtain solutions for 4-dimensional galileons via Kaluza-Klein reduction of Lovelock black hole solutions.

In the next section, we will provide analytic solutions for non shift symmetric galileon theories. Then, in the following section, we will discuss the case of shift symmetric theories in great detail: the no-hair theorem and the two possible known ways to construct black hole solutions. We will develop in particular the surprising features related to time dependence for the galileon field. We will also discuss stability, slow rotation and coupling to matter. We will close with a brief section concerning first results on neutron stars before concluding. We point out to the interested reader that recently, several interesting reviews \cite{Herdeiro:2015waa,Brihaye:2016lin,Volkov:2016ehx,Silva:2016smx} have appeared discussing no-hair theorems and black hole solutions each from a different perspective.

\section{Horndeski black holes from Kaluza-Klein reduction of Lovelock theory}

It has been known since a long time that Lovelock theory{\footnote{For a brief discussion on Lovelock theory in relation to Horndeski theory, see \cite{aegean}.}} is related to scalar-tensor theory via Kaluza Klein reduction \cite{chahoyssen}. The first explicit solutions to Horndeski gravity theory were obtained from black hole solutions of Lovelock theory \cite{blaise}. Let us very briefly review these solutions.  
Consider a $D$-dimensional Einstein Gauss-Bonnet theory which is the full 5 or 6-dimensional Lovelock theory,
\be
	S = \frac1{16\pi G_N}\int d^{D}x\,\sqrt{-g}\left[-2\Lambda +R+ \alpha \hat G \right]. 
\label{chaGBAction2}
\ee
Take the simplest but consistent diagonal reduction along some arbitrary $n$-dimensional internal curved space $\tilde{\mathbf  K}$. We reduce this theory down to $4$ spacetime dimensions with $D=4+n$:
\be
	d s^2_{(4+n)}=d \bar s^2_{(4)} + e^{\phi} d \tilde K^2_{(n)}
	\label{chaKKgalileon}\,.
\ee
This frame is not the most general, it is chosen in such a way as so  there is no conformal factor of $\phi$ in front of the $4$-dimensional metric. As such the radial fall-off of any Lovelock $D$-dimensional black hole solution will be similar to the obtained one  in 4 dimensions. Hence we can expect that the radial fall off in any such truncated solution is akin to higher dimensional black hole solutions; in other words more rapid than the standard GR $1/r$ fall-off. Terms with a tilde refer to the curved $n$-dimensional internal space, while terms with a bar refer to $4$-dimensional spacetime. The idea is now to take the above action and insert the metric ansatz (\ref{chaKKgalileon}), integrate out the $n$-dimensional geometry and obtain a reduced action in 4 dimensions with a scalar field $\phi$ and a 4-dimensional metric. This is possible because the above ansatz (\ref{chaKKgalileon}) is a consistent truncation \cite{blaise}. The important result is that the integer $n$, that corresponds to the dimension of the locally compact Kaluza-Klein space, can be analytically continued to a real parameter of the reduced action \cite{Kanitscheider:2009as}. Therefore $n$ corresponds to a dimension only for $n$ integer. The solutions from the 4-dimensional point of view are still solutions of the resulting effective action for arbitrary real constant $n$. The 4-dimensional effective action reads, after integrating out the internal space,
\ba\label{chaCurvedGBgalileonAction}
%\begin{split}
	&&\bar S_{(4)}= \int d^{4}x\,\sqrt{-\bar g}\,e^{\frac n2\phi}\left\{\bar R -2\Lambda +\frac n4(n-1)\partial\phi^2 -\alpha n(n-1)\bar G^{\mu\nu}\partial_\mu\phi\partial_\nu\phi \right.\nonumber \\
			&+&\alpha \bar G-\frac\alpha4n(n-1)(n-2)\partial\phi^2\nabla^2\phi+\frac\alpha{16}n(n-1)^2(n-2)\left(\partial\phi^2\right)^2 \nonumber\\
			&+&\left.e^{-\phi}\tilde R\left[1+\alpha \bar R+\alpha 4(n-2)(n-3)\partial\phi^2\right]+\alpha\tilde G e^{-2\phi}\right\}\,.
%\end{split}
\label{blaise}
\ea
For $\alpha=0$, this effective action is just the usual KK effective action for (\ref{chaKKgalileon}) with a dilaton and a Liouville type potential $V(\phi)=e^{\frac n2\phi}(-2\Lambda+e^{-\phi}\tilde R)$,  akin to effective string theory actions (see for example \cite{jihad} and references within). Generically, such theories have unusual asymptotics bifurcating no-hair arguments \cite{Chan}.  We note that the higher dimensional Gauss-Bonnet term yields a zoo of galileon terms (associated with the coupling strength $\alpha$). The resulting Horndeski theory is of course not translation invariant  as it is explicitly depending  on $\phi$. Here, we want to obtain a black hole solution of (\ref{chaCurvedGBgalileonAction}) so we need to start from a Lovelock black hole in arbitrary dimension $D$ and then read off the scalar field and the 4-dimensional metric. The important point is to recognize that for locally asymptotically flat spacetime (locally because spacetime may contain a solid angle deficit), the internal metric has to be a product of 2-spheres. This is because we need the 4-dimensional solution to have, at least locally, spherical horizon sections. Thus we have to consider a solution where the $(D-2)$-dimensional horizon sections are $(D-2)/2$-products of two spheres (or in general $(D-2)/2$-products of 2-dimensional constant curvature spaces). 
Thankfully this solution was found in Lovelock theory \cite{dotti} and the transcription in between the two theories, Lovelock (\ref{chaGBAction2}) to Horndeski (\ref{chaCurvedGBgalileonAction}) is a straightforward calculation \cite{blaise}. The solution reads
\ba
	d \bar s^2_{(4)} &= -V(R) d t^2 + \frac{d R^2}{V(R)}+ \frac{R^2}{n+1} d \bar K^2_{(2)}\,, \label{chagalileonMetric}\\
	V(R)&= \kappa+\frac{R^2}{\tar}\left[1\mp\sqrt{1-\frac{2\tar}{l^2}-\frac{2\tar^2 \kappa^2}{(n-1)R^4}+\frac{4 \tar m}{R^{3+n}}}\right], \label{chaGalBHPot}\nonumber\\
	\tar &=  2\alpha n (n+1), \qquad \frac{1}{\ell^2}=\frac{-2\Lambda}{(n+2)(n+3)}, \nonumber\\
	e^\phi&=\frac{R^2}{n+1}\,. \label{chaGalPhi}\nonumber
\ea
Here, $n$ is the dimension of the internal space where we have left out a 2-dimensional constant curvature surface-the 4-dimensional horizon surface; in other words, $n=D-4$. This is the higher dimensional interpretation of $n$ but once the solution is written out, we simply take $n$ an arbitrary real number and (\ref{chagalileonMetric}) is still an exact solution emanating from the theory (\ref{blaise}). In our notation, $\kappa=0,1,-1$ is the normalized horizon curvature with line element $d \bar K^2_{(2)}$ (given in (\ref{constant})) and we have redefined  the constants $\tar$ and $\ell$. Let us take $\kappa=1$, which is the relevant case for positive or zero effective cosmological constant. Taking carefully the $\tar\rightarrow 0$ limit gives a standard Einstein dilaton solution with a Liouville potential \cite{Chan}. These solutions have the following characteristic: the area of the horizon sphere is reduced by $\frac{4\pi R^2}{n+1}$ rather than the standard 2 sphere area $4\pi R^2$ as we can see in (\ref{chagalileonMetric}). This is a solid deficit angle which in the case  $\alpha=0$ yields a central singularity at $R=0$ when there is no mass $m$ to create a horizon. Thus generically these solutions in Einstein-Dilaton gravity are singular --- when there is no black hole mass --- and their asymptotics are akin to the far away field of a global monopole \cite{Barriola:1989hx}. Note the difference with GR, where the $\Lambda=0$ vacuum is Minkowskian  in spherical coordinates (in essence $n=0$), whereas here the wedged sphere vacuum is singular at $R=0$ for any other $n$. The situation for $\alpha\neq 0$ significantly changes as we can see after studying the horizons of the above solution. Note indeed that the vacuum ($m=0$) including the $\alpha$ corrections cloaks the singularity at $R=0$ by producing an event horizon for generic values of the parameter $\alpha$ \cite{blaise}. This solution is a concrete example where higher order terms (part of Horndeski theory) actually cloak a singularity present in a lower derivative action $\alpha=0$. One can notice that the scalar $\phi$ has a logarithmic divergence at spatial infinity, though its energy-momentum tensor remains well-behaved. Also note that the solution parametrized by $n$ interpolates from $n=0$ where we have GR to scalar-tensor solutions with reduced spherical horizon and quicker fall off in the radial direction. The higher order $\alpha$ effects introduce generically an extra internal horizon in the black hole geometry (details can be found in \cite{blaise}). 

\section{Horndeski black holes in shift symmetric theories}
\label{Sec:Black holes}
We will now turn to  black hole spacetimes of theories with shift symmetry. 
Here we should note that although we are mainly interested in spherical horizon geometries for the 4-dimensional solutions, we shall consider for generality, a constant curvature 2 space with line element
\be
\label{constant}
\tx{d}K^2=\frac{d\chi^2}{1-\kappa \chi^2}+\chi^2 d\phi^2,
\ee
where for $\kappa=1$ we have spherical symmetry, for $\kappa=0$ planar symmetry and for $\kappa=-1$ hyperbolic symmetry. The additional cases with $\kappa=-1,0$ are included here for they appear naturally as black hole horizons for negative (effective or bulk) cosmological constant and for Lifshitz type geometries. Although such geometries do not have an immediate interest in cosmology, we include these cases here for completeness as the parameter $\kappa$ appears simply as some normalized parameter in the equations of motion. 
Additionally we take a locally static spacetime and thus we have
\be
\tx{d} s^2 = -h(r) \tx{d} t^2 + \dfrac{\tx{d} r^2}{f(r)} + r^2 \tx{d}K^2.
\label{tansatz}
\ee
The crucial point to note here is that since the scalar field appears only through its derivatives in the Lagrangian, one a priori needs not impose staticity for the scalar. In fact shift symmetric galileons naturally inherit some time dependence \cite{Babichev:2010kj,Babichev:2011iz} in cosmological settings, which is translated to a space and time dependence in a spherically symmetric setting (\ref{tansatz}). This is also true for self-tuning solutions \cite{aegean} as we will see later on in this section (see equation (\ref{st})).  However, this dependence on time cannot be arbitrary. Indeed, in order to have a well defined system of field equations, the 2 tensor that is associated to the variation of the galileon terms with respect to the metric must be static and spherically symmetric. In other words, the associated energy momentum tensor of the galileon must obey the symmetries of spacetime, but not the galileon itself!{\footnote{The same guiding principle is used in GR with a complex scalar field in order to construct a hairy "Kerr" type solution by Herdeiro and Radu \cite{Herdeiro:2014goa}.}}

Treating the general case is possible but technically very tedious, so we will choose to concentrate on specific sub-theories for which one can get analytic results.
So let us concentrate on a subset shift symmetric galileon theory notably,
\be
\mc{L}^\tx{\Lambda CGJ} = R - \eta (\partial \phi)^2 + \beta G^{\mu \nu} \partial_\mu \phi \partial_\nu \phi - 2 \Lambda . 
\label{john}
\ee
This Lagrangian can be obtained by choosing $G_4 = 1 + \beta X$ and $G_2 = -2 \Lambda+ 2 \eta X$. Although the coupling $\eta$ is canonically normalized to $\frac12$, we keep it as $\eta$ momentarily for bookkeeping purposes.   
The field equations are
\bea\label{eomg}
\cal{E}_{\mu\nu}&= G_{\mu\nu} -\eta \left[\partial_\mu\phi \partial_\nu\phi -\frac12 g_{\mu\nu}(\partial\phi)^2 \right]  +g_{\mu\nu}\Lambda \nonumber\\
	&+\frac{\beta}2 \left[ (\partial\phi)^2G_{\mu\nu} + 2 P_{\mu\alpha\nu\beta} \nabla^\alpha\phi \nabla^\beta\phi \right. 
	 \left.   +  g_{\mu\alpha}\delta^{\alpha\rho\sigma}_{\nu\gamma\delta}\nabla^\gamma\nabla_\rho\phi \nabla^\delta\nabla_\sigma\phi \right]
	=0,\nonumber
\eea
The variation of the action with respect to $\phi$ yields
\begin{equation*}\label{eomJ}
 \nabla_\mu J^\mu =0,\;\; J^\mu = \left( \eta g^{\mu\nu} -\beta G^{\mu\nu} \right) \partial_\nu\phi.
\end{equation*}
Here, the key term in the action is the John term from Fab 4 which has nice integrability properties, as we will see. Although our discussion will be associated to the specific action (\ref{john}), the essential results go through quite generically. Sometimes integrability has to be sacrificed on the way in the sense that one has to use numerical methods to obtain solutions.

The effective energy momentum tensor associated to the galileon is precisely ${\cal{T}}_{\mu\nu}=-{\cal{E}}_{\mu\nu}+ G_{\mu\nu}+g_{\mu\nu}\Lambda$. As we noted above, this tensor must obey the symmetries of (\ref{tansatz}) but not the scalar itself. Note for example that the Einstein plus cosmological constant term do not contribute to the ${\cal{T}}_{tr}=0$ equation but other terms in ${\cal{E}}_{tr}$ do. This equation generically describes the inflow of matter in a black hole geometry and will inevitably constrain drastically the galileon field if it is not static. The first key result is the following: 
\begin{framed}
\noindent
Consider the shift symmetric theory (\ref{john}) with spacetime symmetry given by (\ref{tansatz}). Starting with $\phi=\phi(t,r)$ the only compatible ansatz with the field equations is
\be
\phi=q t+ \psi(r).
\label{phiansatz}
\ee
\end{framed}
\noindent
Indeed, taking $\phi=\phi(t,r)$, the flow equation ${\cal{E}}_{tr}=0$  yields the general solution for $\phi$ as a separable function of $t$ and $r$~\cite{Babichev:2013cya}. This function, when inserted in the remaining field equations, gives (\ref{phiansatz}) as the only possible ansatz (see the general discussion in \cite{Appleby}).

The only solution to escape the rule of linear time dependence imposed in (\ref{phiansatz}) is to consider self-tuning solutions for flat spacetime. For theory (\ref{john}), this holds in the case of $\eta=0$ and $\Lambda\neq 0$ . This is  a simple example of a time dependent scalar field immersed in a static spacetime. Indeed, the solution reads
 \be
 \label{st}
 \phi=\phi_0+\phi_1 (r^2-t^2)
 \ee
  with $\phi_0, \phi_1$ integration constants while $f=h=1$ with $\kappa=1$ for (\ref{tansatz}). The self-tuning condition reads $ V_\mathrm{John} \phi_1^2=\Lambda$ for arbitrary bulk $\Lambda$, and constant $V_\mathrm{John}$ \cite{aegean,Appleby}{\footnote{Note that the same solution in a cosmological coordinate system is a purely time dependent function, $\phi=\phi_0+\phi_1 T^2$, where $T$ is FRW proper time. This solution illustrates what we mentioned earlier, a time dependent galileon yields generically a time and space dependent galileon in a static ansatz.}}.

We expect the linear time ansatz (\ref{phiansatz}) to be true for generic shift symmetric theories (the discussion in \cite{Appleby} includes the Paul term; solution (\ref{st}) is also valid for this term, see \cite{aegean}). It is surprising and highly non trivial that there exist time dependent configurations for a static spacetime. Mathematically, we can understand that if time dependence is linear in $t$, we get explicitly ODE's rather than PDE's  once we input (\ref{phiansatz}) in the field equations. It is worthwhile however to make a remark on the non-trivial physical significance of~(\ref{tansatz}) and~(\ref{phiansatz}). 

The metric equation ${\cal E}_{tr}=0$ implies zero influx onto a static metric. Naively, one expects that a time dependent galileon field would accrete onto a black hole, thus making it impossible to keep a static configuration for the metric. Indeed this is the case for  asymptotically flat Fab 4 self-tuning black holes. They are not known analytically but we do know that they cannot be static spacetimes. This is because they are simply not compatible with the (flat) self-tuning solution asymptotics for $\phi$ \cite{Appleby} as one would need $\phi\sim r^2-t^2$ for large $r$. Furthermore, standard or phantom scalar field~\cite{Babichev:2004yx} or even k-essence (which is a particular case of galileon with $G_3=G_4=G_5=0$)~\cite{Babichev:2006vx} with the time-dependent ansatz~(\ref{phiansatz}) 
do accrete onto black holes, rendering a non-zero inflow (or outflow) on a black hole, see e.g. a review~\cite{Babichev:2014lda}. 
Finally, it has been found in~\cite{Babichev:2010kj} that, in the test fluid approximation, galileons also allow accreting solutions, i.e., solutions with a non-static metric. Clearly the solutions which we will describe below belong to a different branch, which exists thanks to the non-trivial higher order structure of galileons. 

Summarizing, starting from $\phi=\phi(t,r)$  and (\ref{tansatz}), we  end up with $\phi = q t+ \psi(r)$ and (\ref{tansatz}) as a starting ground in our search for scalar-tensor black hole solutions where $q$ is a (possibly vanishing) constant. This time dependence was first implemented in \cite{Babichev:2013cya} in order to find the general solution for spacetime (\ref{tansatz}). We will elaborate on this aspect in paragraph \ref{tdep}. 
The results for theory (\ref{john}) with (\ref{phiansatz}) were nicely extended to the framework of the 
shift symmetric Lagrangian \cite{Kobayashi:2014eva} $\mc{L} = G_2(X) + G_4 (X) R + G_{4 X} [(\Box \phi)^2 - (\nabla_\mu \nabla_\nu \phi)^2]$ without significant differences with respect to \cite{Babichev:2013cya}. The method for the more general theory is identical to (\ref{john}) and we refer the reader to the original paper \cite{Kobayashi:2014eva}. Furthermore, for theories without reflection symmetry $\phi \rightarrow - \phi$, in particular for the case of the DGP-like Lagrangian $G_3\neq 0$, the method can still be adapted although it requires numerical integration~\cite{bhkgb}. 

The above represents a very general class of solutions for galileon theories, some of which can be obtained analytically or otherwise numerically. A second important and distinct class involves the Gauss-Bonnet term (\ref{GB}) which is a topological invariant in 4 dimensions. This means that if this term is not coupled to the galileon field $\phi$ in the action, it yields no term in the field equations. Indeed, this stems from the first Lovelock identity, valid only in $4$ spacetime dimensions:
\be
\label{chaddual}
H_{\mu\nu}=-2 P_{\mu cde}R_\nu{ }^{cde}+\frac{{g_{\mu\nu} }}{2}\hat G=0 \, ,
\ee
which basically tells us that the metric variation of (\ref{GB}) is trivial in 4 dimensions.
Here we have noted $H_{\mu\nu}$ the Lovelock 2-tensor obtained from the metric variation of the Gauss-Bonnet term (\ref{GB}). When this term couples linearly to the scalar field as $\phi \hat G$, it recovers shift symmetry.  As we will see, such a term gives, whenever present, a distinct class of numerical solutions with $q=0$ \cite{Sotiriou:2014pfa}.
But first we shall start by establishing a no-hair theorem with its precise hypotheses, which will in turn indicate the possible ways to obtain non trivial hairy solutions.

\subsection{A no-hair theorem}

Hui and Nicolis were the first to consider and point out no-hair arguments for shift symmetric galileon theories \cite{Hui:2012qt}. A specific way out to their argument was discussed in \cite{Babichev:2013cya}, where an explicit solution method was found and generic solutions were given. Sotiriou and Zhou looked in greater detail to the no-go theorem and developed the arguments of Hui and Nicolis while providing another class of numerical solutions~\cite{Sotiriou:2013qea,Sotiriou:2014pfa}. 
A straightforward generalization to the arguments of Hui and Nicolis was developed by Maselli \textit{et al.} \cite{Maselli:2015yva} to extend to cases of linear time dependence (\ref{phiansatz}) but, as we will see in the next section, this case nicely bifurcates the no-hair theorem by use of the field equations. Thus, given all these works and considerations, let us first concentrate on the static $q=0$ case giving the following no-go theorem.

\begin{framed}
\noindent Consider a shift symmetric galileon theory as (\ref{ssgg})  where $G_2$, $G_3$, $G_4$, $G_5$ are arbitrary functions of $X$. We now suppose that:
\begin{enumerate}
\item spacetime is spherically symmetric and static (\ref{tansatz}) while the scalar field is also static ($q=0$),
\item spacetime is asymptotically flat, $\phi'\rightarrow 0$ as $r\rightarrow \infty$ and the norm of the current $J^2$ is finite on the horizon,
\item there is a canonical kinetic term $X$ in the action and the $G_i$ functions are such that their $X$-derivatives contain only positive or zero powers of $X$.
\end{enumerate} 
Under these hypotheses, we conclude that $\phi$ is constant and thus the only black hole solution is locally isometric to Schwarzschild.  
\end{framed}

Indeed, using the symmetry assumptions, it is useful here to rewrite the line element (\ref{tansatz}) as
\begin{equation*}
\tx{d} s^2 = -A(r) \tx{d} t^2 + \dfrac{\tx{d} r^2}{A(r)} + \rho(r)^2 (\tx{d} \theta^2 + \sin^2\!\theta\, \tx{d} \varphi^2).
\label{metrichh}
\end{equation*}
The norm of the current is $J^\mu J_\mu =(J^r)^2/A$. By assumption, the norm of the current does not diverge on the horizon. Hence, when we are the horizon location $A=0$, $J^r$ can only vanish. The conservation equation now gives $\nabla_\mu J^\mu = \rho^{-2} (\rho^2 J^r)' =0$ which implies that $\rho^2 J^r$ is constant. The quantity $\rho$ is the areal radius, used to measure the area of constant radius spheres. The latter should not be singular (zero or infinite), even at the horizon. This means that $\rho^2 J^r$ vanishes at the horizon and hence it vanishes everywhere. $J^r$ is therefore zero everywhere.
Now $J^r$ can be put under the form $J^r = A \phi' F(\phi';g,g',g'')$, where the explicit expression of $F$ is given in \cite{Sotiriou:2014pfa}. At any point, either $F$ or $\phi'$ has to vanish. Under assumption (iii) that was made on the Lagrangian and because of asymptotic flatness, $F \rightarrow - G_{2X} = constant$ as $r\rightarrow +\infty$. Hence $\phi'$ is zero everywhere. Then the only solution is locally isometric to the GR solution. We should emphasize that the physical hypothesis in this theorem is that the norm of the current is finite  as it is associated to the shift symmetry of the Lagrangian. 

\subsection{The no-hair theorem and hairy black holes} 

\begin{figure}[t]
\begin{center}
\includegraphics[width=\textwidth]{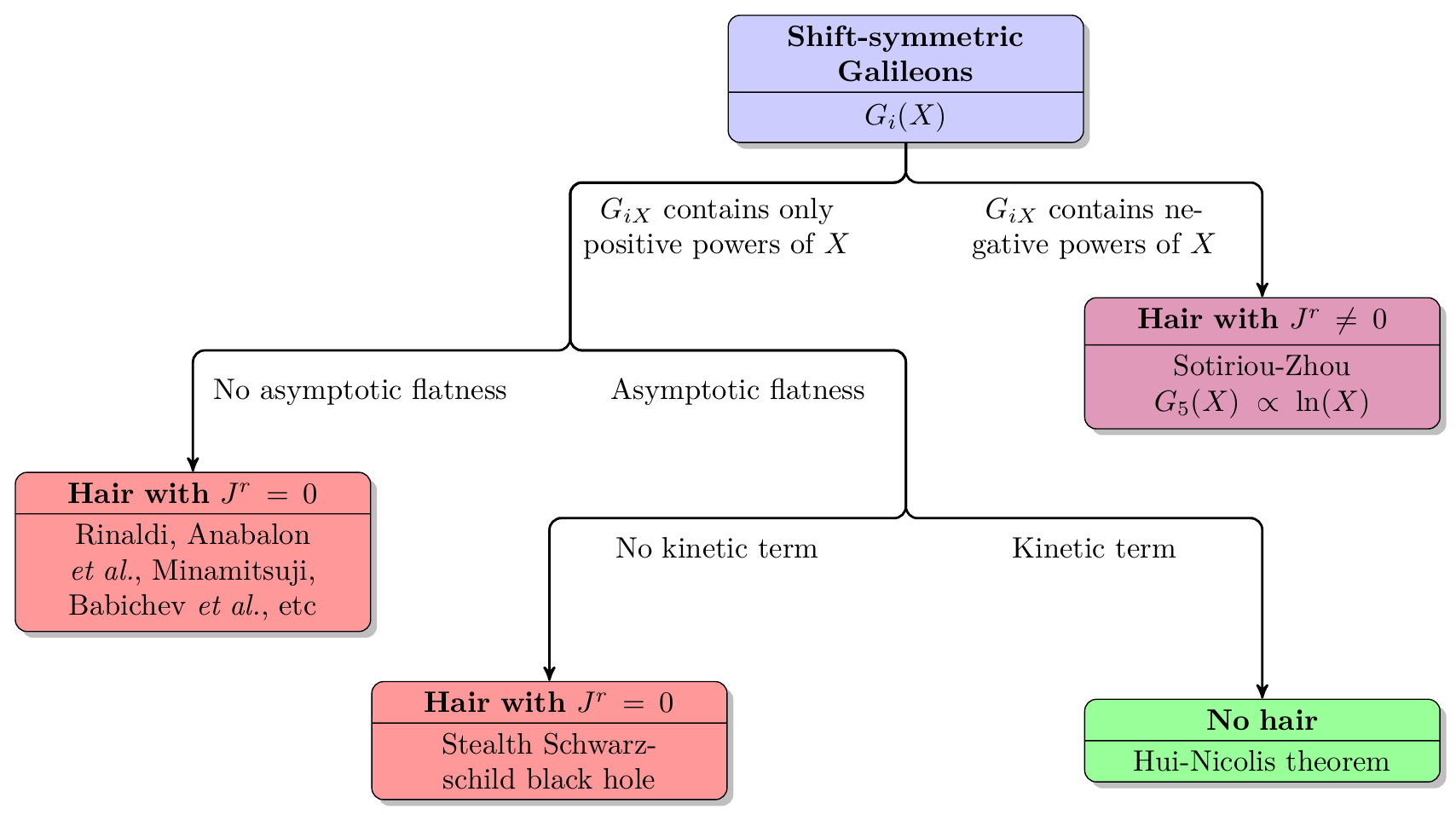}
\caption{Hair versus no-hair}
\label{valid}
\end{center}
\end{figure}

There actually exist (at least) two ways to construct black hole solutions with a non-trivial scalar field, summarized in Fig. \ref{valid}. The first of these is to have such a theory as to be able to set everywhere $J^r=0$ without imposing that $\phi'=0$. Generically in this case we will either give up asymptotic flatness or the presence of a canonical kinetic term. The second is to include a Gauss-Bonnet term in the action coupled to a linear scalar field. We now look at these methods in turn. 

The first family amounts to necessarily taking higher derivative terms which allows $F=0$ without $\phi'=0$ in the notation of the above theorem. This case is naturally realized for the time-dependent ansatz (\ref{phiansatz}). The key result is the following  \cite{Babichev:2015rva}:

\begin{framed}
\noindent
Consider an arbitrary shift symmetric Horndeski (or beyond Horndeski) theory and a scalar-metric ansatz dictated by (\ref{tansatz}), (\ref{phiansatz}) with $q\neq 0$. The only solution to the scalar field equation ${\cal E}_\phi=0$ and the ``matter flow'' metric equation ${\cal E}_{t r}=0$ is  given by $J^r=0$.
\end{framed}
\noindent
Indeed, as demonstrated in \cite{Babichev:2015rva} we have that:
\begin{equation*}
-q J^r={\cal E}_{t r} f,
\end{equation*}
and given that ${\cal E}_\phi=\nabla_\mu J^\mu=0$, the result trivially follows.  
The current now reads
\begin{equation*}
J^\mu J_\mu =-A (J^t)^2+(J^r)^2/A,
\end{equation*}
and $J^t$ can even be singular like $1/\sqrt{A}$ while the current is regular on the horizon. We emphasize that the physical requirement of the no-hair theorem above is now satisfied by virtue of the field equations.

If the theory is of higher order there will be solutions other than the trivial case $\phi'=0$ as we will see in a moment. In fact requiring that $\phi'=0$ and $q\neq 0$, in the case of (\ref{john}), always leads to singular solutions as was shown very recently in \cite{Charmousis:2015txa}. Although a general proof for an arbitrary shift symmetric theory is not known, we expect it to remain true. Under the assumption that $q\neq 0$, the field equations dictate regularity of the current and indicate the presence only of non-trivial scalar field solutions. 

On the other hand we note that the integration constant associated to the scalar field equation is equal to zero since $J^r=0$. Hence, the would be ``primary charge" is set to zero whenever time dependence is present and is replaced by the velocity parameter~$q$. 

If $q=0$, then we have to go back to the no-hair argument and the regularity of $J$ in order to set $J^r=0$ by hand. 

Therefore, we see that imposing time dependence immediately renders the no-hair theorem irrelevant, and a higher order Horndeski theory such as (\ref{john}) immediately imposes $J^r=0$ with $\phi'\neq 0$. Furthermore, $J^r=0$  simultaneously annihilates two of the field equations and gives a mathematically consistent system of field equations as for three variables $f$, $h$, $\psi$ there are three remaining independent field equations: $J^r=0$, ${\cal E}_{r r}=0$ and ${\cal E}_{t t}=0$. Therefore non trivial solutions to the field equations a priori exist. A last important remark here is that time dependence renders the scalar field regular in the future black hole horizon, something which is never true for $q=0$ (for details see \cite{Babichev:2013cya}). Again, a proof of this is only known for theory (\ref{john}), but we expect it to remain true in general. We will summarize the open problems for the reader in the last section. 

Our discussion above has been quite general; the integrability requirement applies even to beyond Horndeski theories \cite{Babichev:2015qma}. To make the points stressed above clearer let us consider our example theory, (\ref{john}).
Here, $J^r= \phi' (\eta g^{rr}-\beta G^{rr})$ and  $F=\eta g^{rr}-\beta G^{rr}$ in the notation of the theorem above. So we can have either $\phi=constant$ and $q=0$ corresponding to the usual GR black holes with trivial scalar field, or $\eta g^{rr}-\beta G^{rr}=0$. The case $\phi'=0$ and $q\neq 0$ has been ruled out in  \cite{Charmousis:2015txa}.

The second family of solutions involves the Gauss-Bonnet invariant $\hat G$.
The no-hair theorem assumes that the $X$-derivatives $G_{iX}$ contain only positive or zero powers of $X$. But now suppose that  the $G_{iX}$  contain terms that are $G_5 \propto \mathrm{ln} |X|$ and therefore $G_{5X} \propto 1/X$. In fact, this along with $G_i=0$ for $i=2,3,4$ is the term $\alpha \phi \hat G$ involving the Gauss-Bonnet invariant (\ref{GB}). Since this is a topological term in 4 spacetime dimensions it does not contribute to the metric field equations with Lovelock tensor $H_{\mu\nu}=0$ (\ref{chaddual}). Such a Lagrangian still has shift-symmetry, because $\hat G$ is locally a total divergence from Poincar\'e's lemma.  For simplicity, consider a Lagrangian built of the usual Ricci scalar, a minimal kinetic term and the Gauss-Bonnet term:
\be
\label{gblag}
\mc{L}^\tx{GB}=   \frac{R}{2} - \frac12 \pd_\mu \phi \pd^\mu \phi +\alpha \phi \hat G.
\ee
One gets the following scalar equation:
\begin{equation*}
\Box \phi + \alpha \hat G = 0 .
\end{equation*}
Therefore, here it is essential that we keep the minimal kinetic term. Now this theory because of the Gauss-Bonnet term does not satisfy requirement iii) of the theorem. Furthermore, the Gauss-Bonnet term becomes a geometric source of d'Alembert's equation for $\phi$ on a curved background. It becomes necessary that $\phi$ is non trivial in such a setup in a very natural way. Sotiriou and Zhou insist that any theory where this Gauss-Bonnet term is not forbidden \textit{must} have hairy black holes \cite{Sotiriou:2013qea,Sotiriou:2014pfa}. They go on to find a particular solution with $q=0$ to the above theory (\ref{gblag}) using numerical and perturbative methods.{\footnote{The perturbative solution was first obtained in a remarkable paper involving a charged rotating black hole dressed with an axion and a dilaton by Campbell \textit{et al.} \cite{kaloper}. Numerical solutions of a very similar theory ($\phi \mc{G}\rightarrow e^\phi \mc{G}$) were found early on in Mavromatos \textit{et al.} \cite{Kanti:1995vq}.}}. They found that there exists a finite radius singularity which can be hidden behind the horizon if the black hole is massive enough. They also remarked that corrections to observables in this model are expected to be small with respect to GR.

A relation in between the scalar charge and the mass of the black hole is fixed in order for the scalar to be regular on the event horizon. The black hole constructed is then with secondary hair. This solution is therefore clearly separate from the previous class for two reasons: it acquires $q=0$ and, more importantly, a non zero scalar charge (to be fixed relative to the black hole mass). Therefore $J^2$ is actually singular at the horizon because $J^r=-f \phi' -4 \alpha \frac{h'}{h}\frac{f(f-1)}{r^2}\neq 0$. 
 At this point one needs to invoke extra input to conclude about the physical relevance of solutions with divergent norm of the current $J$. 
For this solution of the theory~(\ref{gblag}), the Noether current cannot be a physical observable,
in particular, it cannot be coupled to matter directly. This question requires further study.

\subsection{ Explicit solutions of hairy black holes}
\label{tdep}

We shall now concentrate on explicit black hole solutions for the theory (\ref{john}) setting $\eta=\frac12$. Although the method works for any shift symmetric theory, the advantage here is that (\ref{john}) is particularly elegant in giving explicit solutions. In fact, we have the general solution which we turn to now. 
\begin{framed}
\noindent
The general solution of theory (\ref{john}) to the metric (\ref{tansatz}) and $\phi=\phi(t,r)$ is given as a solution to the following third order algebraic equation with respect to $\sqrt{k(r)}$:
\bea
\label{master}
&(q\beta)^2 \left(\kappa +\frac{r^2}{2\beta}\right)^2 -  \left(2\kappa+( 1-2\beta \Lambda ) \frac{r^2}{2\beta}\right) k(r) + C_0 k^{3/2}(r) =0, 
\eea
\end{framed}
\noindent
where $C_0, q$ are integration constants and $\kappa=1,-1,0$ is the horizon curvature. Once a solution to the above algebraic equation is given, the metric components are
\be
\label{h}
h(r) = -\frac{\mu}{r} +\frac{1}{\beta r}\int \frac{k(r)}{\kappa+\frac{r^2}{2\beta}}dr, \qquad f = \frac{ (\kappa+ \frac{r^2}{2\beta})^2 \beta h}{ k(r)},
\ee
whereas the scalar field (\ref{phiansatz}) reads
\begin{equation*}
\psi' = \pm \frac{\sqrt{r}}{h(\kappa+\frac{r^2}{2\beta})}\left(q^2 (\kappa+ \frac{r^2}{2\beta}) h'-\frac{1+2\beta\Lambda}{4\beta^2}(h^2r^2)'\right)^{1/2}.
\end{equation*}
An explicit proof can be found in \cite{Babichev:2013cya} where here for the master equation we have rescaled $C_0$ by $\beta$ and set $\eta=\frac12$ with respect to \cite{Babichev:2013cya}. We note that the algebraic equation is parametrized by $q\beta$, $\beta\Lambda$ and $C_0$ and the overall sign of $\beta$. We now work out the different classes of solutions according to their asymptotic behavior for large $r$.

\subsubsection{Class I: dS and adS asymptotics}

For dS or adS asymptotics it is easy to see from (\ref{master}) that $k=\alpha r^4=\frac{1-2\beta \Lambda}{2\beta C_0} r^4$ as $r\rightarrow \infty$. Since we want $f=h$ for $r\rightarrow \infty$ we get that $C_0=\frac{1-2\beta \Lambda}{\sqrt{\beta}}$. Therefore, once we fix $C_0$ to this value, keeping $q$ arbitrary, we have asymptotically de Sitter or anti de Sitter solutions. The generic solution is found as a solution to the algebraic solution, but a simple example is the self-tuned Schwarzschild-(anti-)de Sitter spacetime which is given by
\be
\label{desitter}
k_0(r)=\left(\kappa+\frac{r^2}{2\beta}\right)^2, 
\ee
where now the parameter $q_0$ is fixed:
\be
\label{q0}
q_0^2 \beta=(1+2\beta \Lambda)\kappa
.\ee
For de Sitter asymptotics, we take $\kappa=1$ and the solution reads
\be
\label{self}
f=h=1- \dfrac{\mu}{r}- \dfrac{\Lambda_\tx{eff}}{3} r^2, ~~~~\psi'= \pm \dfrac{q}{h} \sqrt{1-h}
,\ee

\noindent where the effective cosmological constant $\Lambda_\tx{eff}=-\frac{1}{2\beta}$ and hence $\beta<0$ for $\Lambda_\tx{eff}>0$. This solution has self-tuning properties since the vacuum value of $\Lambda$ does not interfere with the spacetime geometry. It is  tuned by the integration constant $q_0$ via (\ref{q0}). It is quite remarkable that this self-tuning solution, first noted for pure de Sitter \cite{Linder}, can be extended for generic black holes \cite{Babichev:2013cya}. A characteristic of this particular solution is that the kinetic scalar $X=q_0^2/2$ is a constant. 

This self-tuning solution of de Sitter is therefore very special since $q_0$ (and not only $C_0$) is fixed with respect to the parameters of the action (\ref{q0}). But in fact we will now argue that self tuning remains beyond this particular value, $q=q_0$, where of course $X$ is not constant.  This would mean that a change in the bulk cosmological constant will not change the self-tuning mechanism, in other words the effective cosmological constant remains the same. To see this, suppose that $q^2 \beta=  (1+2\beta \Lambda)+\epsilon$, where $\epsilon$ is some small number compared to  $1+2\beta \Lambda$.  We now consider an expansion in $\epsilon$ to $k=k_0+\epsilon k_1$ where $k_0$ is given in (\ref{desitter}). It is then easy from (\ref{master}) to show that
\begin{equation*}
k(r)=\beta \left(1+\frac{r^2}{2\beta}\right)^2\left( 1+ \frac{2\epsilon}{1+6\beta\Lambda-(1-2\beta \Lambda)\frac{r^2}{2\beta}}\right)+O(\epsilon^2).
\end{equation*}
Evaluating $f$ and $h$, we see that the black hole solution remains unchanged asymptotically and in particular that the effective cosmological constant is not modified to this order in $\epsilon$ \cite{damos}. In other words, the self-tuning mechanism remains true even if the bare cosmological constant changes. The novel $q=q_0+\epsilon$ tunes to accommodate the new value of the bulk cosmological constant, while the black hole solution is different locally but has the same asymptotic behavior. 

This class of solutions also includes the ``static" $q=0$ solutions. 
This condition was first imposed by Rinaldi~\cite{Rinaldi:2012vy} to obtain exact solutions in asymptotically adS spacetimes, although the scalar field could become imaginary beyond the black hole horizon. Indeed, Eq.(8) of \cite{Rinaldi:2012vy} implies imaginary scalar field inside the horizon.  Rinaldi's solution was extended in \cite{Anabalon:2013oea,Minamitsuji:2013ura} who cured the problem by including a bare cosmological constant. The scalar field was found to be divergent at the horizon even though the norm was finite. At the end of the day, this is not necessarily a problem, since at the level of the action, only derivatives of the galileon field itself are present and on-shell the action is well behaved.
Indeed, from (\ref{master}) we obtain:
$$
k(r)= \frac{1}{C_0^2}\left(2\kappa+( 1-2\beta \Lambda ) \frac{r^2}{2\beta}\right)^2.
$$
Executing the integral (\ref{h}) we can evaluate directly $h$. We set for convenience $\kappa=1$, i.e., spherical symmetry and $\beta>0$. We also fix $C_0$ accordingly, in order to avoid a solid deficit angle. In other words, we set the constant term in $h$ to be equal to $1$. We then get the ``static" solution, first discovered in \cite{Rinaldi:2012vy} for $\Lambda=0$ and extended in \cite{Babichev:2013cya,Anabalon:2013oea,Minamitsuji:2013ura}:
\begin{equation}
\label{phirsol}
\eqalign{ 
h(r) = 1- \dfrac{\mu}{r} - \dfrac{\Lambda_\mathrm{eff}}{3 }  r^2 - \dfrac{(1+2 \beta \Lambda_\mathrm{eff})^2}{8 \beta \Lambda_\mathrm{eff}} \: \dfrac{\tx{Arctan} \left(r/\sqrt{2\beta}\right)}{r/\sqrt{2\beta}}, \\
f(r) = \dfrac{(2 \beta + r^2) h}{2 \beta (r h)'}, \quad \phi'^2 = \dfrac{(1+2\beta \Lambda_\mathrm{eff})r^2(1-2 \beta \Lambda_\mathrm{eff} -2 \Lambda_\mathrm{eff} r^2)^2}{\Lambda_\mathrm{eff} (1- 2 \beta \Lambda_\mathrm{eff}) (2 \beta +r^2)^3 h(r)},
}
\end{equation}
with an effective cosmological constant $\Lambda_\mathrm{eff} =\frac{2\beta \Lambda-1}{2 \beta (2\beta\Lambda+3)}$.

\subsubsection{Class II: Static Universe}
From (\ref{master}) we see that $k\sim   r^2$ as $r \rightarrow \infty$ for $q^2\beta= (1-2\beta \Lambda)$ while asymptotically $h=1$ and $f=\frac{r^2}{2\beta}$. A typical example in this class of metrics is the black hole embedded in an Einstein static universe, which is obtained with $C_0=0$ and, for simplicity, $2\beta \Lambda=-1$. The solution reads
\begin{equation}\label{fh2}
	h = 1- \frac{\mu}{r}, \;\;  f=\left(1- \frac{\mu}{r}\right)\left(1+\frac{\eta r^2}{\beta}\right),
\end{equation}
whereas the radial part of the scalar field is given by
\begin{equation*}\label{psi2}
\psi' = \pm \frac{q}{h}\sqrt{\frac{\mu}{r (1+\frac\eta\beta r^2) }}
\end{equation*}

An alternative way to obtain explicit solutions can be given in this class \cite{damos}.  Let us start with the de Sitter solution (\ref{desitter})  $k_0$ in Class II. Consider the Euclidean division of the third order polynomial in (\ref{master}) by $\sqrt{k}-\sqrt{k_0}$ which is a factor of the third order polynomial. Obviously, the resulting Euclidean division gives a second order polynomial in $\sqrt{k}$ from which it is easy to read off the two remaining possible solutions. This technique allows us to obtain additional solutions (for the same set of parameters) once we have an explicit root of the polynomial. As our example shows the remaining roots of the polynomial generically belong to different classes of solutions. In other words, the asymptotic behavior is generically different. Additionally, for the solutions to remain real, one may have to consider different signs of the fixed parameters. For the explicit solution, see \cite{damos}.

\subsubsection{Class III: anisotropic scaling and Lifschitz spacetimes}
The last possibility in the presence of a cosmological constant and a canonical kinetic term is to have $k\sim r^{8/3}$ while $2\beta \Lambda=1$. In this case, it is easy to see that as $r\rightarrow \infty$, we have $h(r)\sim r^{2/3}$ and $f\rightarrow r^2$. These spaces have a Lifschitz spacetime behavior and their scaling symmetry is anisotropic. Indeed, we have
$$
t\rightarrow \lambda^z t,\quad r\rightarrow \frac{r}{\lambda}, \quad (x,y)\rightarrow \lambda(x,y),
$$
and for our case here the Lifschitz exponent is $z=1/3$. Lifschitz 4-dimensional black holes are finite temperature duals of 3-dimensional non relativistic scale invariant theories in the context of holography. In this case, it is more interesting to take flat horizon asymptotics, $\kappa=0$ \cite{Bravo-Gaete:2013dca}, \cite{Ayon-Beato:2015qfa}. The planar black hole solution from (\ref{master}) was found by Bravo-Gaete and Hassaine in \cite{Bravo-Gaete:2013dca} and has geometry
$$
ds^2=-r^{2/3} g(r) dt^2+\frac{dr^2}{r^2 g(r)}+r^2 (dx^2+dy^2),
$$
with $g(r)=1-\frac{m}{r^{5/3}}$. 

\subsubsection{Class IV: No canonical kinetic term }
Setting $\eta=0$ in the action (\ref{john}) and apart from the flat self-tuning solution (\ref{st}), we obtain the following algebraic equation to solve,
$$
(q\beta)^2  -  (2- \Lambda r^2) k(r) + C_0 k^{3/2}(r) =0, 
$$
where we have set $\kappa=1$ for simplicity. The metric components are,
$$
h(r) = -\frac{\mu}{r} +\frac{1}{\beta r}\int k(r) dr, \qquad f(r)= \frac{  \beta h}{ k(r)},
$$
whereas the scalar field (\ref{phiansatz}) reads
$$
\psi' = \pm \frac{\sqrt{r}}{h}\left(q^2  h'-\frac{\Lambda}{2\beta}(h^2r^2)'\right)^{1/2}.
$$
Again there are different type of solutions in particular taking $q=0$ we get a black hole solution which asymptotically behaves as $h\sim r^4$ and is valid only for $\Lambda<0$.  
Another example solution can be obtained by taking $C_0=0$. We then get,
$k(r)=\frac{q^2 \beta^2}{2-\Lambda r^2}$.

Let us terminate with asymptotically flat spacetimes where $\Lambda=\eta=0$. 
We obtain a unique solution,
for $k=constant$:
\begin{equation}\label{fh1}
 f  =h = 1- \frac{\mu}{r},
\end{equation}
and the metric is isometric to a Schwarzschild metric (BTZ stealth black holes were found in 3 dimensional spacetimes \cite{AyonBeato:2004ig}). However, the scalar field is not trivial and also regular in the future black hole horizon \cite{Babichev:2013cya}, $\psi'=\pm q \sqrt{\mu r}/(r-\mu)$. The fact that we take $\Lambda=\eta=0$ may arguably lead to strong coupling in flat spacetime ($\mu=0$) for the scalar field. Note however that the black hole solution that is found is identical to GR, so strong coupling does not a priori create a phenomenological problem, as local gravity tests remain indistinguishable relative to GR. 

Furthermore, the stealth Schwarzschild metric has also the property $X=q^2/2$ (kinetic term is constant on-shell).  Because of this property it is easy to show that the above stealth solution remains a solution of the beyond Horndeski theory:
\begin{equation*}
\mc{L}^\tx{bH} = R + F_\tx{J}(X) G^{\mu \nu} \partial_\mu \phi \partial_\nu \phi ,
\label{Bjohn}
\end{equation*}
where $F_\tx{J}(X)$ is a function of $X$ only \cite{Babichev:2015qma}. It is unknown as yet if other $X=contant$ solutions can be extended in a similar way to beyond Horndeski theories.

\subsection{Stability}

As we discussed in the beginning, Horndeski theory avoids Ostrogradski ghosts, because the field equations remain second order, and new degrees of freedom are not present. It is however not clear if the existing propagating degrees of freedom --- the scalar spin-0 and the tensor spin-2 --- are healthy degrees of freedom for each particular model. 
Moreover, there are indications that galileon theory contains a {\it nonlinear} ghost instability (which can be interpreted as a {\it globally} unbounded from below Hamiltonian), 
see e.g. a discussion in~\cite{Sivanesan:2011kw}. 
This however is not an issue as such, since there may exist a local minimum with a long-lived vacuum state.
It is therefore more important to check if relevant solutions for a particular model at hand form a {\it locally} stable vacuum.  
For this it is convenient to use a perturbative approach, i.e., one studies whether small perturbations around a specific solution are stable or not. 
There may exist different types of pathologies, including ghost, gradient or tachyon 
instability\footnote{On the nonlinear level yet another pathology may manifest itself: 
formation of caustics, which is generic for theories with nonlinear $G_2$ as a function of $X$~\cite{Babichev:2016hys}.}.

The question of stability of black hole solutions in Horndeski theory has not been fully investigated up to now, although some works have been dedicated to the topic. 
In particular, Kobayashi {\it et al.}~\cite{Kobayashi:2012kh,Kobayashi:2014wsa} focused on the stability of general spherically symmetric black holes with static galileon field,
using the Regge-Wheeler formalism~\cite{Regge:1957td}. 
Necessary conditions were established to ensure absence of ghost and gradient instabilities.
Tachyon  instability has been left out in this study.
Particular subclasses of Horndeski theory were also considered in~\cite{Kobayashi:2012kh,Kobayashi:2014wsa}, including the John term and the Gauss-Bonnet term coupled to galileon. It was shown that the static John solution (\ref{phirsol}) is stable in some range of the $\beta$, $\Lambda$ parameters. In the small coupling limit, the Gauss-Bonnet black hole of Ref. \cite{Sotiriou:2014pfa} also appears to be stable, quite naturally.
Later, Cisterna \textit{et al.}~\cite{Cisterna:2015uya} derived the odd parity perturbation equation for the model (\ref{john}) plus a potential term $V(\phi)$. 
In this sector of Horndeski theory, they showed the mode stability under linear odd-parity perturbations of black holes with static galileon.
This study should be considered as complementary to~\cite{Kobayashi:2012kh,Kobayashi:2014wsa}. 
On one hand, it does not allow to identify ghosts if present, while on the other hand it allows to conclude about the presence of tachyons.

Although the study of static galileon solutions is an important task, which can give us an insight about the behavior of black holes in Horndeski theory,
it is not clear if such a setup is physically relevant. As we have already mentioned in the beginning of this section, 
shift symmetric galileons naturally inherit some time dependence due to the cosmological evolution. 
Therefore, the study of time-dependent solutions seems to be a more physically relevant task.

Potentially viable  black hole solutions should recover homogeneous cosmology at large distances. 
Alternatively, switching off the mass of a black hole, one should also find a cosmological solution. 
The study of perturbations in a homogeneous universe is normally simpler than the corresponding analysis for a black hole that is immersed
in such a universe, thus offering an alternative check of the instability of solutions. Of course, the stability of a homogeneous solution does not necessarily mean the stability of 
a black hole solution, but any pathology found for a homogeneous solution means ruling out a corresponding black hole solution with this homogeneous asymptotic. 
A particular example, for which we can use the results on cosmological perturbations analysis
is the black hole in the self-tuned de Sitter~\cite{Babichev:2013cya}.
Taking the Lagrangian~(\ref{john}) and assuming vanishing black hole mass, 
we can apply the results of Appleby and Linder~\cite{Appleby:2011aa} (for zero $\Lambda$). 
It turns out that in such expanding universe, the scalar perturbations suffer from a gradient instability~\cite{Appleby:2011aa}.

Recently, Ogawa {\it et al.} established an important instability result~\cite{Ogawa:2015pea}. They studied odd-parity mode perturbations of the time-dependent black hole solutions found in~\cite{Babichev:2013cya,Kobayashi:2014eva}. The considered theory included galileon terms $G_2$ and $G_4$, but not $G_3$ or $G_5$, thus the theory possesses reflection symmetry $\phi \rightarrow -\phi$ in addition to shift symmetry. Their analysis is made at the level of an effective action and does not involve the boundary conditions which are however essential for the galileon field since it is time and space dependent.  
For odd perturbations, the second order perturbation Lagrangian $\mc{L}^{(2)}$ contains only one dynamical degree of freedom, let us call it $\chi$. 
For the ansatz~(\ref{tansatz}), one finds~\cite{Ogawa:2015pea}
\be
\label{Lpert}
\mc{L}^{(2)} = c_1 \: \dot{\chi}^2 - c_2 \: \chi'^2 + c_3 \: \dot{\chi} \chi' + c_4 \: \chi^2,
\ee
where $c_i$ are functions of the background solution. In order for the corresponding Hamiltonian to be positive, $c_1$ and $c_2$ have to be positive. However, Ref.~\cite{Ogawa:2015pea} found that for solutions such that $X=constant$, the product $(c_1 c_2)$ becomes negative in the vicinity of a horizon.
Thus, it is claimed that an instability is present in the odd-parity modes near an event or cosmological horizon for any solution such that $X=constant$. The resulting instablility for the particular self tuned black hole (where $q=q_0$ is fixed) agrees with the cosmological instability found in~\cite{Appleby:2011aa} (although for the black hole it is not clear which type of instability is present).
For a particular choice of the Lagrangian, namely when $G_4 = 2 \: X G_{4 X}$, the perturbation Lagrangian~(\ref{Lpert}) vanishes and the theory becomes strongly coupled. This subspace of solutions concerns the black hole in an Einstein static universe found in \cite{Babichev:2013cya}. 
A similar instability is also present in higher dimensions, as shown in~\cite{Takahashi:2015pad} in the framework of so called Lovelock-galileon theory.

\subsection{Slowly rotating black holes}
\label{slowrotsec}

Given the general solution for spherically symmetric and static black holes, one can question what happens for rotating black holes i.e., in the case of stationary rather than static spacetimes. To this date, no exact solutions have been found to this non-linear problem. It seems that integrability present for spherical symmetry ceases once we allow for a stationary metric as the scalar field would be also angle-dependent. One can look for linear corrections to the found solutions, caused by slow rotation. 
A standard way to proceed is to adopt the Hartle-Thorne perturbative approach of GR \cite{Hartle:1967he,Hartle:1968si}, where the metric is taken in the following form:
$$
\tx{d} s^2 = -h(r) \tx{d} t^2 + \dfrac{\tx{d} r^2}{f(r)} + r^2 (\tx{d}\theta^2 + \tx{sin}^2 \theta \tx{d}\varphi^2) - 2 \omega(r) r^2 \tx{sin}^2 \theta \tx{d}t \tx{d}\varphi,
%\label{slowrot}
$$
and the frame-dragging function $\omega(r)$ is of linear order in the angular velocity of the black hole. 
The scalar field is assumed to have the same form as in the non-rotating case~(\ref{phiansatz}), where the parameter $q$ is zero for the Sotiriou and Zhou solution \cite{Sotiriou:2013qea}.
Then the metric and scalar equations are solved at first order in $\omega(r)$, giving a second order ODE on $\omega(r)$.
In GR, the following equation emerges:
\begin{equation}\label{HTeq}
	\omega''+\frac{\omega'}{2} \left(\frac{f'}{f}+\frac{8}{r}-\frac{h'}{h}\right)=0.
\end{equation}
In modified gravity, one expects that (\ref{HTeq}) is changed. 
Indeed, Maselli \textit{et al.}~\cite{Maselli:2015yva} (see also the early work of \cite{kaloper})
found that for the linear (and also exponential) coupling of the scalar to Gauss-Bonnet term, the equation describing 
the correction to the non-rotating solution of Ref.~\cite{Sotiriou:2013qea} is modified with respect to~(\ref{HTeq}).

Furthermore, for the example theory (\ref{john}), the same Ref.~\cite{Maselli:2015yva} found that the equation on the frame-dragging function is modified in the following way:
\be
\label{strong}
2 (1 - \beta X) \left[\omega''+\frac{\omega'}{2} \left(\frac{f'}{f}+\frac{8}{r}-\frac{h'}{h}\right)\right] - 2 \beta X' \omega' =0.
\ee
For a generic solution, $X'$ will not vanish and the frame-dragging function will differ from GR. Yet, for particular solutions of \cite{Babichev:2013cya} like stealth (\ref{fh1}) and self-tuning (\ref{self}) black holes, $X'$ does vanish and the equation on $\omega$ is unchanged with respect to GR. The Einstein static universe black hole (\ref{fh2}) both satisfies $X'=0$ and $(1 - \beta X)=0$, therefore (\ref{strong}) disappears signaling strong coupling in accordance to \cite{Ogawa:2015pea}. Thus, we see that solutions close or identical to GR pick up identical corrections for the frame-dragging function to linear order. But we emphasize that it is not so for the generic solution of (\ref{master}), where clearly an important deviation from GR occurs due to $X'\neq 0$.

We note finally that (\ref{HTeq}) can be integrated once and given in terms of the $k=k(r)$ function that parametrizes the general solution (\ref{master}):
$$
(1 - \beta X) \omega'= \frac{C_1 \sqrt{k}}{r^4 (1+\frac{r^2}{2\beta})} 
,$$
with $C_1$ an arbitrary integration constant.

\subsection{Coupling to matter fields and to additional galileons}

A natural question now arises : how do the above solutions extend in the presence of matter? 
In particular, how will matter couple to the above system? Consider for a start a $U(1)$ gauge field in the following action.
\bea
\mc{L}^\tx{EM} = R - \eta (\partial \phi)^2 + \beta G_{\mu \nu} \partial^\mu \phi \partial^\nu \phi - 2 \Lambda - \dfrac{1}{4} F^{\mu \nu} F_{\mu \nu} - \gamma T_{\mu \nu}^\tx{EM} \pd^\mu \phi \pd^\nu \phi ,
\nn
T_{\mu \nu}^\tx{EM} = \dfrac12 \left(F_{\mu \rho} F_\nu^{\; \rho} - \dfrac14 g_{\mu \nu} F^{\rho \sigma} F_{\rho \sigma}\right)
\label{john+EM}\nonumber
.\eea
Note that apart from the usual Maxwell electromagnetic term, we have a galileon-EM interaction term, dressed with a coupling $\gamma$ \cite{Babichev:2015rva}. It may seem dangerous to include such a higher derivative coupling between the scalar and the matter sector but let us pursue for the moment. 
The scalar field equation is now given by
\begin{equation*}
\nabla_\mu J^\mu =0,
\end{equation*}
where
\begin{equation*}\label{J}
    J^\mu = \left(\beta \, G^{\mu\nu}-\eta\,g^{\mu\nu}-\gamma\,T^{\mu \nu}_{(M)}\right) \nabla_{\nu}\phi.
\end{equation*}
Thus, in a similar spirit as before, the scalar field equation or more precisely the current components reproduce one of the metric Einstein Maxwell equations. This was anticipated to play an important role in rendering the theory integrable. 
Taking the following dyonic type ansatz:
\bea 
\tx{d} s^2 = -h(r) \tx{d} t^2 + \dfrac{\tx{d} r^2}{f(r)} + r^2 \tx{d}\Omega^2, ~~~~\phi(t,r) = q t + \psi(r),
\nn
A_\mu \tx{d}x^\mu = A(r) \tx{d}t + B(\theta) \tx{d}\varphi,\quad \phi=qt+\psi(r)
\label{EMansatz}\nonumber
,\eea
the field equations reduce  to an algebraic equation of fifth order this time \cite{Babichev:2015rva}. Explicit solutions can be found in the original paper. The interesting general fact is that they are closer to GR solutions when $\gamma\neq 0$. In particular, we can obtain the extension of the de Sitter self-tuning black hole, which reads:
\bea\label{solfh}
h(r)= 1-{\frac {\mu}{r}} +{\frac {\eta\,{r}^{2}}{3\,\beta}}+{\frac { \gamma\, \left( {Q}^{2}+{P}^{2} \right) }{4\,\beta\,{r}^{2}}}, \nonumber\\
(\psi'(r))^2 = \frac{1-f(r)}{f(r)^{2}} q^2,\nonumber\\
F_{tr} =F(r)=\frac{Q}{r^{2}}, \quad F_{\theta \varphi}
=C(\theta)=P\,\sin(\theta),\nonumber
\eea
with
\begin{equation*}\label{constfh}
{P}^{2}\beta\, \left( \Lambda\,\gamma+\eta \right) ={Q}^{2}\eta\,
 \left( \gamma-\beta \right)
,\quad q^2=\frac{\eta + \Lambda\,\beta}{\beta\,\eta},\quad
C_{0}=\frac{1}{\eta} \left(\eta-\beta\,\Lambda\right).
\end{equation*}
The above solution is a galileon version of the Reissner-Nordstr\"om (RN) black hole with a cosmological constant which, unlike its GR counterpart,  persists into hiding the vacuum cosmological constant from the spacetime geometry. Furthermore, we see that despite the non-minimal coupling of the Maxwell field, the solution is
identical to the dyonic version of the Reissner-Nordstr\"om-de-Sitter
metric, up to redefinition of $Q$ and $P$. One can verify
that the metric, the gauge field and the norm of the Noether current
are regular at the horizon. The scalar field is regular at the
future black hole horizon.
Solutions were also found for $q=0$ and $\gamma=0$ by Cisterna and Erices \cite{Cisterna:2014nua}, asymptotically behaving like a (a)dS metric and a Coulomb potential, as well as an uncommon ``electric universe'' where the metric is locally flat and the electric field goes to a finite value at spatial infinity. See also \cite{Kolyvaris:2011fk}. 

In the same frame of mind as above, we now consider the coupling of an additional scalar field. We start with the following theory including an Einstein-Hilbert term with a conformally coupled scalar field: 
\be
\label{bek}
{\cal{L}}^\tx{BBMB} = R +  \eta \left[ -\dfrac12 (\pd \phi)^2 - \dfrac{1}{12} \phi^2 R \right]
.\ee
The scalar field equation is invariant under the conformal transformation
$$
 g_{\alpha\beta}\rightarrow \tilde{g}_{\alpha\beta} = \Omega^2 g_{\alpha\beta},\quad
\phi \rightarrow  \tilde{\phi} = \Omega^{-1} \phi .
$$
This is the gravitational action of the BBMB solution. The solution has the geometry of an extremal black hole:
$$
d s^2 = -  (1-\frac{m}{r})^2 d t^2+ \frac{d r^2}{( 1-\frac{m}{r} )^2} + r^2 (d\theta^2 + \sin^2\theta d\varphi^2 ) ,
$$
with secondary scalar hair:
$$
\phi = \sqrt{\frac{3}{4\pi G}} \frac{m}{r-m},
$$
since the only free parameter is the black hole mass, $m$. We see that there is no independent charge for the scalar field. The latter explodes at the horizon location. Can we remedy this with a time dependent galileon field? To this end we now consider adding to the action a galileon term:
\begin{equation*}
{\cal{L}}^\tx{Biscalar} = {\cal{L}}^\tx{BBMB} +(\beta G_{\mu \nu} - \gamma T^\tx{Conf}_{\mu \nu}) \pd^\mu \Psi \pd^\nu \Psi,
\label{biscal}
\end{equation*}
where 
$$
T^\tx{Conf}_{\mu\nu}=\frac{1}{2}\nabla_{\mu}\phi\nabla_{\nu}\phi-\frac{1}{4}g_{\mu\nu}\nabla_{\alpha}\phi\nabla^{\alpha}\phi
+\frac{1}{12}\left(g_{\mu\nu}\square-\nabla_{\mu}\nabla_{\nu}+G_{\mu\nu}
\right)\phi^{2}~. 
$$
Again, note the non trivial coupling of the galileon field $\Psi$ with the scalar $\phi$. As before, the action has been engineered so that the $\Psi$-variation gives
\begin{equation*}
\label{psii}
\nabla_{\mu}J^{\mu}=0~,
\qquad J^{\mu}=\left(\beta G^{\mu\nu}-\gamma T^{\mu\nu}
\right)\nabla_{\nu}\Psi~.
\end{equation*}
The important point here is that the current vector
$J^\mu$ ``contains'' the metric field equations of the BBMB action (\ref{bek}). This turns out to be crucial in finding the spherically symmetric solution. We consider
$$
ds^{2}=-h(r)dt^{2}+\frac{dr^{2}}{f(r)}+r^{2}d\Omega^{2}~, 
$$
with
$$
\phi=\phi(r), \,\,\,\,\,\, \Psi=q t+\psi(r)~.
$$
The solution reads \cite{Charmousis:2014zaa}
\bea
&&f(r)=h(r)=1-\frac{m}{r}+\frac{\gamma c_{0}^{2}}{12\beta r^{2}}~,\nonumber\\
&&\phi(r)=\frac{c_{0}}{r}~,\nonumber\\
&&\psi'(r)=\pm q \frac{\sqrt{1-h}}{h},\nonumber\\
&& \beta=\frac{2}{3q^2},\quad\gamma=\frac{2}{q^2}~. \nonumber
\eea 
Note that the solution{\footnote{We thank T. Kolyvaris for communicating a typo in the original version of the solution appearing in \cite{Charmousis:2014zaa}.}} has the following nice properties. For a start, the scalar $\phi$ now has a genuine scalar charge $c_0$ that is independent of the mass parameter $m$. It is also singular at $r=0$, as is the curvature tensor, but not on the horizon location. The metric is identical to a RN metric, except that it is now sourced by the two scalars.

\section{Neutron stars}

So far we have been focusing on black hole solutions in galileon theory i.e., solutions that have an event horizon. 
It is natural to ask what happens if a smooth matter source is placed instead of a black hole, so that the solution now describes a star configuration.

It turns out that the stealth solution (\ref{fh1}) of Ref.~\cite{Babichev:2013cya} can be naturally modified to include a smooth source.
Indeed, Cisterna {\it et al.}~\cite{Cisterna:2015yla} considered Lagrangian~(\ref{john}) with $\eta =\Lambda=0$ and a matter source that does not couple to the galileon. 
For this model, there exists a vacuum ``exterior" solution with stealth Schwarzschild metric (\ref{fh1}), Ref.~\cite{Babichev:2013cya}. 
Thus, one can match an interior solution inside the star to the exterior stealth solution. 
In order to correctly match the exterior stealth and the interior solutions, the construction of an interior solution should follow the same procedure as the exterior solution of~\cite{Babichev:2013cya}. 
Namely, the radial component of the Noether current must be set to zero, $J^r=0$, as in the case of the vacuum solution.
Then, the only modification with respect to the vacuum equations is the presence of density and pressure of matter.

The equation $J^r=0$ leads however to a rather non-trivial consequence in the presence of a source term. Indeed, 
this equation directly relates the two metric functions, $f(r)$ and $h(r)$ of the metric~(\ref{tansatz}). 
In the presence of matter, this relation does not coincide with the relation between $f_\tx{GR}$ and $h_\tx{GR}$ found in GR. 
Therefore, inside matter, the metric is not the same as in GR. 
Indeed, Cisterna {\it et al.}~\cite{Cisterna:2015yla} numerically integrated the field equations and found neutron star solutions with some deviation from GR solutions.
For a given radius, $\beta<0$ implies heavier neutron stars, while $\beta>0$ implies lighter stars compared to GR.
An interesting consequence of this work is that in the limit $q\rightarrow 0$ (while keeping $\beta= constant$),
the GR solution is not recovered (although the solutions appear to be relatively close).

The emission of gravitational waves by shift-symmetric Hornedski neutron stars was studied in \cite{Barausse:2015wia}. Assuming the validity of a Post-Newtonian (PN) expansion and a static galileon, it was concluded that no difference should be expected with respect to GR at first PN orders. 

It is interesting to note here a relation to the Vainshtein mechanism~\cite{Arkady}, which normally operates in galileon theories (see e.g. a review~\cite{Babichev:2013usa}).
The solution described above does not give any modification to GR outside the source.
This is in contrast to a ``usual'' Vainshtein mechanism, when GR is restored inside some radius, with small but nonzero corrections.

Cisterna \textit{et al.} extended their work to more realistic equations of state and slowly rotating solutions in \cite{Cisterna:2016vdx}. 
Their study takes into account most neutron stars, except millisecond pulsars. 
They require the theory to be able to reproduce the heaviest known pulsar, PSR J0348+0432, which has mass $2.01 \pm 0.04 \; M_\odot$ \cite{Antoniadis:2013pzd}. 
When $\beta > 0$, this yields a maximal value for $\beta q^2$ (see \cite{Cisterna:2016vdx} for the details and numerical values). When $\beta < 0$, there is apparently no constraint from this requirement. Gamma ray bursts from the pulsar SGR0526-066 provide an additional test for redshift \cite{Higdon:1990dv}. The authors checked that their solutions are consistent with the allowed redshifts $z = 0.23 \pm 0.07$ for masses between 1 and $1.5 M_\odot$. The redshift analysis does not provide sharper constraints than the maximal mass test, though. 
References~\cite{Cisterna:2016vdx} and \cite{Silva:2016smx} also initiated a study of the relation between mass and moment of inertia for neutron stars. Maselli \textit{et al.} conducted a similar study in~\cite{Maselli:2016gxk}. Combining their results with the stability analysis of \cite{Ogawa:2015pea}, they found that sensibly compact stars should lead to exterior solutions that are stable under odd-parity perturbations.

%%%%%%%%%%%%%%%%%%%%%%%%%%%%%%%%%%%%%%%%%%%%%%%%%%%%%%%%%%%%%%%%%%%%%%%%%%%%%%%%%%%%%%%%%%%%%%%%%%%%%%%%%

\section{Conclusions}

We have discussed black hole and star solutions in the context of higher order scalar-tensor theories. We first saw how these can be constructed from Lovelock theory using standard Kaluza Klein methods. These are the only solutions known analytically that do not possess shift symmetry in the scalar field. The solutions have non trivial asymptotics and are generically found to shield with an inner horizon a solid angle deficit angle singularity. 

The case of shift symmetric theories is far less usual or standard, and possesses solutions which have very intriguing properties. There are two generic types of solutions found: ones that involve a Gauss-Bonnet term with a linear galileon coupling, or, ones that naturally involve linear time dependence in the galileon field and for which the radial current component is $J^r=0$. For the former, the Gauss-Bonnet term acts as a source to the scalar field equation \cite{Sotiriou:2013qea}, immediately disallowing a trivial scalar field in curved spacetime. These solutions, as we explained, have a diverging current on the horizon. The current is associated to shift symmetry of the action and thus may be a physical characteristic of the black hole. An interesting open problem is to study Gauss-Bonnet black holes with a time dependent galileon. 

The latter are naturally associated to a time-dependent galileon. The field equations have additional branches of solutions relative to GR, due to the higher order nature of galileons, and this is of key importance in obtaining hairy black holes. For the theory (\ref{john}), the general solution is known explicitly \cite{Babichev:2013cya}, whereas integrability has also been shown for the shift symmetric theory with time reflection symmetry \cite{Kobayashi:2014eva}. We saw that given time dependence for the galileon,  the field equations bifurcate the no-hair theorem while satisfying stronger regularity conditions than those imposed by the theorem for the scalar field. We saw that there exist solutions that are GR-like while having self-tuning properties for the bulk cosmological constant. In other words, the bulk cosmological constant does not fix the cosmological constant horizon, it is fixed by the strength of the higher order galileon term. These GR type solutions are strongly correlated to the Lagrangian John of Fab 4. On the negative side, these solutions, with kinetic energy $X=constant$ for the galileon, appear to have some pathology as shown very recently in \cite{Ogawa:2015pea}. Is this instability related to the case $X=constant$ and what does this case mean geometrically? It is clear that $X=constant$ solutions are very special and simplify the relevant expressions but we should emphasize that they are not the generic solution to the field equations. We saw in particular that self-tuning properties clearly go beyond the simple self-tuning solution.

Furthermore, it is probable that demanding spacetime to be static may be the reason for instability. It is already surprising that static metric solutions exist with a time dependent galileon, since we know that Birkhoff's theorem is not true for scalar-tensor theories.  Further research in this direction is necessary. Another issue, associated to star metrics, is how the galileon couples to standard matter inside the star.  Up to now, the hypothesis has been that the galileon does not couple to matter. We saw however that for the case of electromagnetism or bi-scalar black holes, a non-standard coupling can lead to solutions which are far closer to GR solutions.  Clearly there are numerous questions to be treated in this quite novel field involving strong gravity solutions of scalar tensor theories. 

\section*{Acknowledgements}

The authors  acknowledge financial support from the research program, Programme national de cosmologie et Galaxies of the CNRS/INSU, France.
EB was supported in part by Russian Foundation for Basic Research Grant No. RFBR 15-02-05038.

\end{document}